\documentclass[pre,11pt,onecolumn,preprint]{revtex4} 
\usepackage{amssymb}
\usepackage{latexsym}
\usepackage{amsfonts}
\usepackage{graphics}
\usepackage{graphicx}
\usepackage{epsfig}
\usepackage{amsmath}
\usepackage{float}
\usepackage{verbatim}
\usepackage[subfigure]{graphfig}

\begin{document}

\title{Resilience of networks with community structure behaves as if under an external field}

\author{Gaogao Dong,$^{a,b,c}$ Jingfang Fan,$^{d}$ Louis M. Shekhtman,$^{d}$ Saray Shai,$^{f}$ Ruijin Du,$^{a,b,c}$ Lixin Tian,$^{g,b}$ Xiaosong Chen,$^{h,i}$ H.Eugene Stanley$^{c}$ and Shlomo Havlin$^{d,e}$}

\affiliation{$^a$Institute of applied system analysis, Faculty of Science, Jiangsu University, Zhenjiang, 212013 Jiangsu, China.
$^b$Energy Development and Environmental Protection Strategy Research Center, Faculty of Science, Jiangsu University, Zhenjiang, 212013 Jiangsu, China.
$^c$Center for Polymer Studies and Department of Physics, Boston University, Boston, MA 02215, USA.
$^d$Department of Physics, Bar-Ilan University, Ramat-Gan 52900, Israel.
$^e$Tokyo Institute of Technology, Tokyo, Japan.
$^f$Department of Mathematics, University of North Carolina, Chapel Hill, NC 27599, USA.
$^g$School of Mathematical Sciences, Jiangsu Center for Collaborative Innovation in Geographical Information Resource Development and Application, Nanjing Normal University, Jiangsu, 210023. P.R. China
$^h$School of Physical Sciences, University of Chinese Academy of Sciences, Beijing 100049, China.
$^i$CAS Key Laboratory of Theoretical Physics, Institute of Theoretical Physics, Chinese Academy of Sciences, Beijing 100190, China.
}

\date{April 20, 2018}
\begin{abstract}
Detecting and characterizing community structure plays a crucial role in the study of networked systems. However, there is still a lack of understanding of how community structure affects the systems' resilience and stability. Here, we develop a
framework to study the resilience of networks with community structure based on
percolation theory. We find both analytically and numerically that the
interlinks (connections between the communities) affect the percolation phase transition in a manner similar to an external field in a ferromagnetic-paramagnetic  spin system.  We also study the universality class by defining the analogous critical exponents $\delta$ and $\gamma$, and find that their values for various models and in real-world co-authors networks follow fundamental scaling relations as in physical phase transitions. 
The methodology and results presented here not only facilitate the study of resilience of networks but also brings a fresh perspective to the understanding of phase transitions under external fields.

\end{abstract}
\pacs{89.75.Hc, 64.60.ah, 89.75.Fb}
\maketitle

Network science has opened new perspectives in the study of complex systems in social, technologiccal, biological, climate systems and
many other fields \cite{watts_collective_1998,barabasi_emergence_1999,cohen_complex_2010,newman2010networks,boccaletti_complex_2006,fan2017network,boers2014prediction}.
A system's resilience (or robustness) is a key property and plays a crucial role in reducing risks and mitigating damages \cite{cohen_resilience_2000,gao_universal_2016}.
Percolation theory is an effective tool for understanding and evaluating resilience through topological and structural properties \cite{coniglio1982cluster,sokolov1986dimensionalities,coniglio1977percolation,aharony2003introduction,bunde2012fractals}. It is essentially concerned with analyzing the connectivity of components throughout a network. It has been applied to many natural and man-made systems \cite{saberi2015recent}. Critical phenomena in social and complex networks  have attracted researchers from different disciplines \cite{dorogovtsev_critical_2008}. In particular researchers have studied the existence of phase transitions in connectivity (percolation) \cite{cohen_complex_2010,newman2010networks}, the more stringent k-core percolation \cite{dorogovtsevcore_2006,liu_core_2012}, epidemic spreading models \cite{newman_random_2001,brockmann2013hidden,hufnagel2004forecast}, condensation transitions and the Ising model on
complex networks \cite{dorogovtsev_ising_2002}.  It has been pointed out that a random network undergoes a continuous percolation phase transition for
increasing fraction of random node failures \cite{bollobas2001random}. The question of whether discontinuous percolation transitions in networks exist has attracted much attention \cite{achlioptas2009explosive,riordan2011explosive,cho2013avoiding}. Buldyrev \textit{et al.} developed a model of interdependent networks and found analytically that the percolation transition is
discontinuous due to the emergence of cascading failures between the networks. A framework for understanding the robustness of interdependent networks was then developed, and found that a system of interdependent networks undergoes an abrupt
first-order percolation phase transition \cite{buldyrev_catastrophic_2010,gao_networks_2012,yuan2017eradicating,kivela2014multilayer,boccaletti2014structure,shekhtman2016recent,reis2014avoiding,gao2015recent}.

Aside from these significant advancements in understanding the resilience properties of various networks, much work has performed on interconnected networks \cite{leicht2009percolation} such as those formed by connecting several communities (or modules) \cite{wang2013effect,radicchi2013abrupt}. This community structure is ubiquitous in many real-world networks including brain networks \cite{meunier2010modular,stam2010emergence,morone2017model}, infrastructure \cite{guimera2005worldwide,eriksen2003modularity}, social networks \cite{girvan2002community,liu2012social,thiemann2010structure}, and others \cite{lancichinetti2009detecting,onnela_structure_2007,gonzalez2007community,mucha2010community}. Despite these important advances, a central realistic feature of these networks that has not been considered is that usually only a small fraction of nodes are able to sustain inter-module connections. We show here that this feature changes dramatically the resilience of such networks. This small fraction of interconnecting nodes is often due to a need of special resources or infrastructure support. For example, if we consider the airport network, it is known that only some airports have international flights mainly because of the need for longer runways for large planes, customs and passport control, etc. \cite{guimera2005worldwide}. At the same time, once a node has the capability for interconnections, the costs of adding additional such connections are likely to be considerably smaller. Similarly in social networks, it is likely that only certain individuals possess the necessary skills to bridge between different communities \cite{gladwell2006tipping} and in power grids only some power stations have the capacity to supply other far away stations. Here we develop a model to incorporate these realistic features and using methods from statistical physics, provide an analytic solution showing that the interconnections can be described as having effects analogous to an external field in spin systems. This allows us to gain a fundamental understanding of the effects of adding interconnections and make predictions for resilience.

\section{Model}
\begin{figure}
\centering
\includegraphics[width=0.5\linewidth]{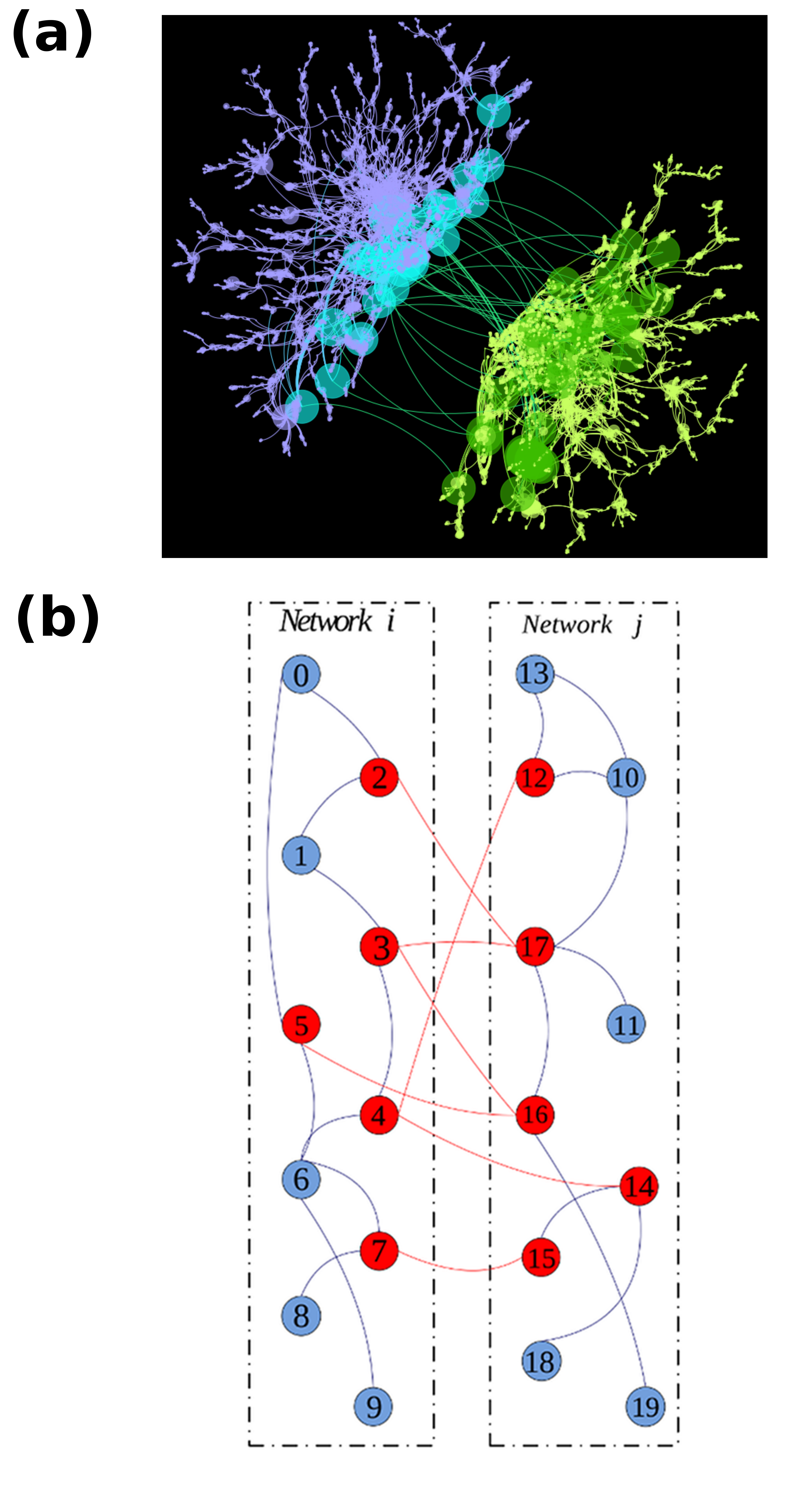}
\caption{\label{Fig:1} (a) Demonstration of two interconnected modules from the co-author collaboration network (dblp). Nodes are authors and a link between two nodes exists if two authors have published at least one paper together. 
(b) Demonstration of the model. We assume two modules $i$ and $j$ and connect a fraction $r$ of nodes according to our model.
Red nodes denote interconnected nodes and the fraction of interconnected nodes here is $r$=1/2.
}
\label{fig:frog}
\end{figure}

Our model is based on the modular structure present in many real-world networks, where a number of well connected groups of nodes (modules) have only some nodes with connections to other modules,  see Fig.~\ref{fig:frog}(a). We show here that these \textit{interconnected nodes} act in a manner analogous to an external field from physics \cite{Stanley_1971,huang2009introduction,reynolds1977ghost}. To study this effect, we demonstrate, for simplicity, a network of two modules, labeled \textbf{i} and \textbf{j}, each with the same number of nodes, \textit{N}. The theory below is general for $m$ communities.
Within module \textbf{i} [\textbf{j}], the nodes are randomly connected with degree distribution $P_{i}(k)$ [$P_{j}(k)$], where the degree, $k$, is defined as the number of links a particular node has to other nodes within module \textbf{i} [\textbf{j}]. Between modules \textbf{i}  and \textbf{j}, we randomly select a fraction, $r$, of nodes as interconnected nodes, and randomly assign $M_{inter}$ interconnected links among pairs of nodes (one in \textbf{i} and the other in \textbf{j}).  A network generated from this model can be seen in Fig.~\ref{fig:frog}(b). The generalization to $m$ modules is obvious.


To quantify the resilience of our model, we study both, analytically and via simulations, the size of the giant connected component $S(r,p)$ after randomly removing a fraction $1 - p$ nodes.

%

\section{Theory}
Here we present a theoretical framework for studying the robustness of our interconnected communities network model.  To obtain an analytic solution, we adopt the generating function framework of \cite{newman2010networks} and define the generating function of the degree distribution for each module $i$ as
\begin{equation}
G_{i}(\textbf{x})=(1-r_i)G_{ii}(x_{ii})+r_iG_{ii}(x_{ii})\prod _{j \neq i}^{m} G_{ij}(x_{ij}), i=1,2, \cdots, m\\
\label{eq1}
\end{equation}
where $G_{ii}(x_{ii})$ and $G_{ij}(x_{ij})$ represent the generating functions of the intra-connections in module $i$ and the inter-connections between modules $i$ and $j$ respectively. We note that $m$ is the number of modules. For the generating functions of the excess degree distribution, see SI \cite{SI}. After randomly removing a fraction $1 - p$ of nodes, the size of the giant component within module $i$ becomes,
\begin{equation}\label{eq2}
S_{i}=p(1-G_{i}(1-p(1-f_{ii}),1-p(1-f_{ij}))),
\end{equation}
where $x_{ii}=1-p(1-f_{ii})$ and $x_{ij}=1-p(1-f_{ij})$.

For the case of $m=2$  Erd\H os-R\'enyi (ER) \cite{erdos1960evolution} modules with average intra-degree $k$, and inter-degree $K$, ~\eqref{eq2} becomes,
\begin{equation}\label{eq3}
e^{-Sk}(r-1)+1-\frac{S}{p}=re^{Kp(\frac{e^{-Sk}(r-1)+1-\frac{S}{p}-r}{r})-Sk}.
\end{equation}
\emph{\textbf{where $K=\frac{2M_{inter}}{rN}$}} and $N$ is total number of nodes in network. For $r = 0$, our model is equivalent to the ER model and we obtain $S=p(1-e^{-kS})$, in agreement with the well known result \cite{erdos1960evolution}.
For $r>0$, the giant component of single ER disappears at $p_{c}=\frac{1}{k}$ and we obtain a percolation threshold for our model that is proportional to $r$, which is assumed to be small \cite{SI}.

We also consider scale-free (SF) modules with power-law degree distribution $P(k)\sim k^{-\lambda}$. The same generating function framework is used to obtain the giant component and percolation threshold  (see SI \cite{SI}).

\section{Results}

\begin{figure}
\centering
\includegraphics[width=.9\linewidth]{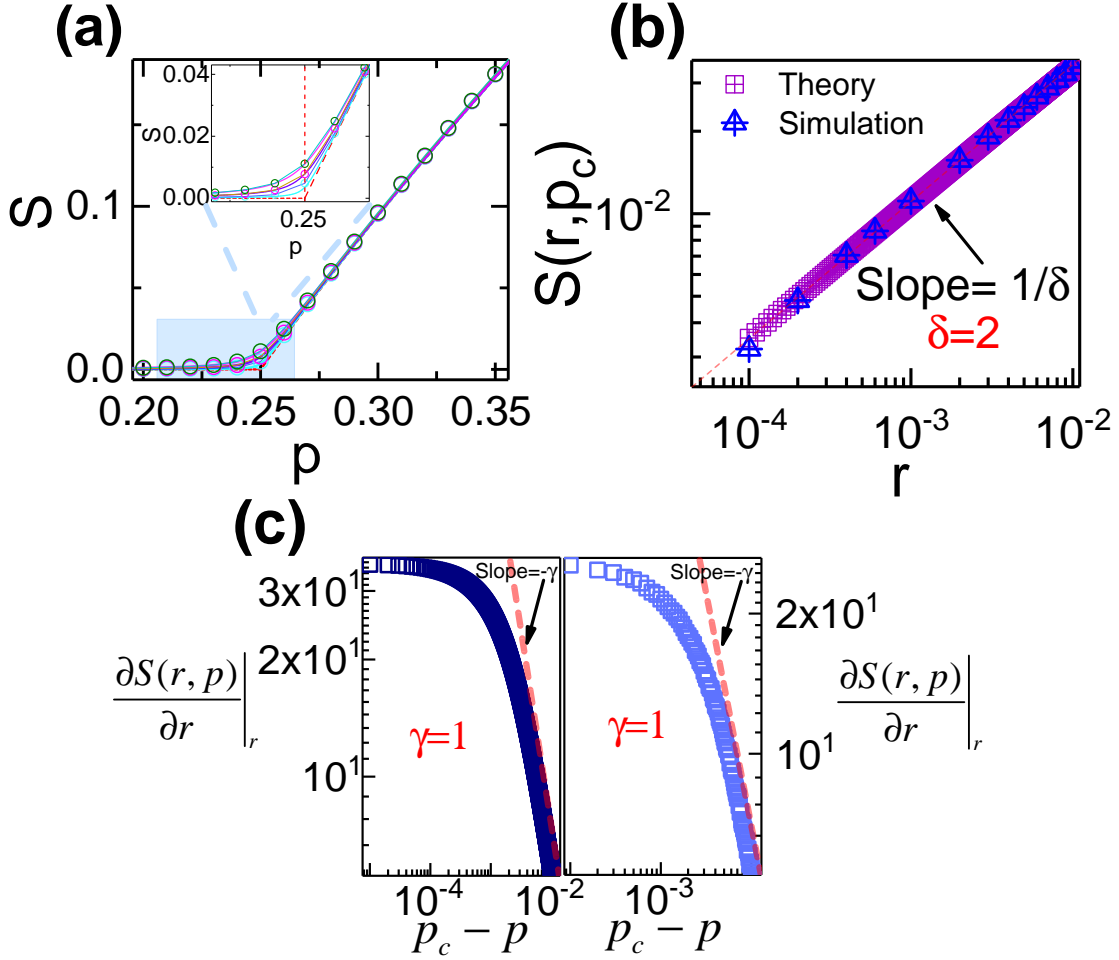}
\caption{\label{Fig:2} (a) Comparison of analytical and simulation results for ER networks for the size of the giant component $S(r,p)$ as a function of $p$ with $r=0$ (red), $r=0.0001$ (blue), $r=0.0055$ (purple) and $r=0.001$ (magenta). Lines and symbols denote analytical and simulation results respectively.
(b) $S(r,p_c)$ as a function of $r$. (c) $\frac{\partial S(r,p)}{\partial r}$ as a function of $p_{c}-p$ with $r=0.0001$. Left and right panels show the numerical and simulation results respectively.
The parameters are $k=4$, $M_{inter}=N_{1}$ and for simulation results we chose the size of modules to be $N_{1}=N_{2}=10^{8}$, $M_{inter}=N_{1}$ and averaged over 1000 realizations. Similar results for different parameters are given in SI.}
\end{figure}

\begin{figure}
\centering
\includegraphics[width=.9\linewidth]{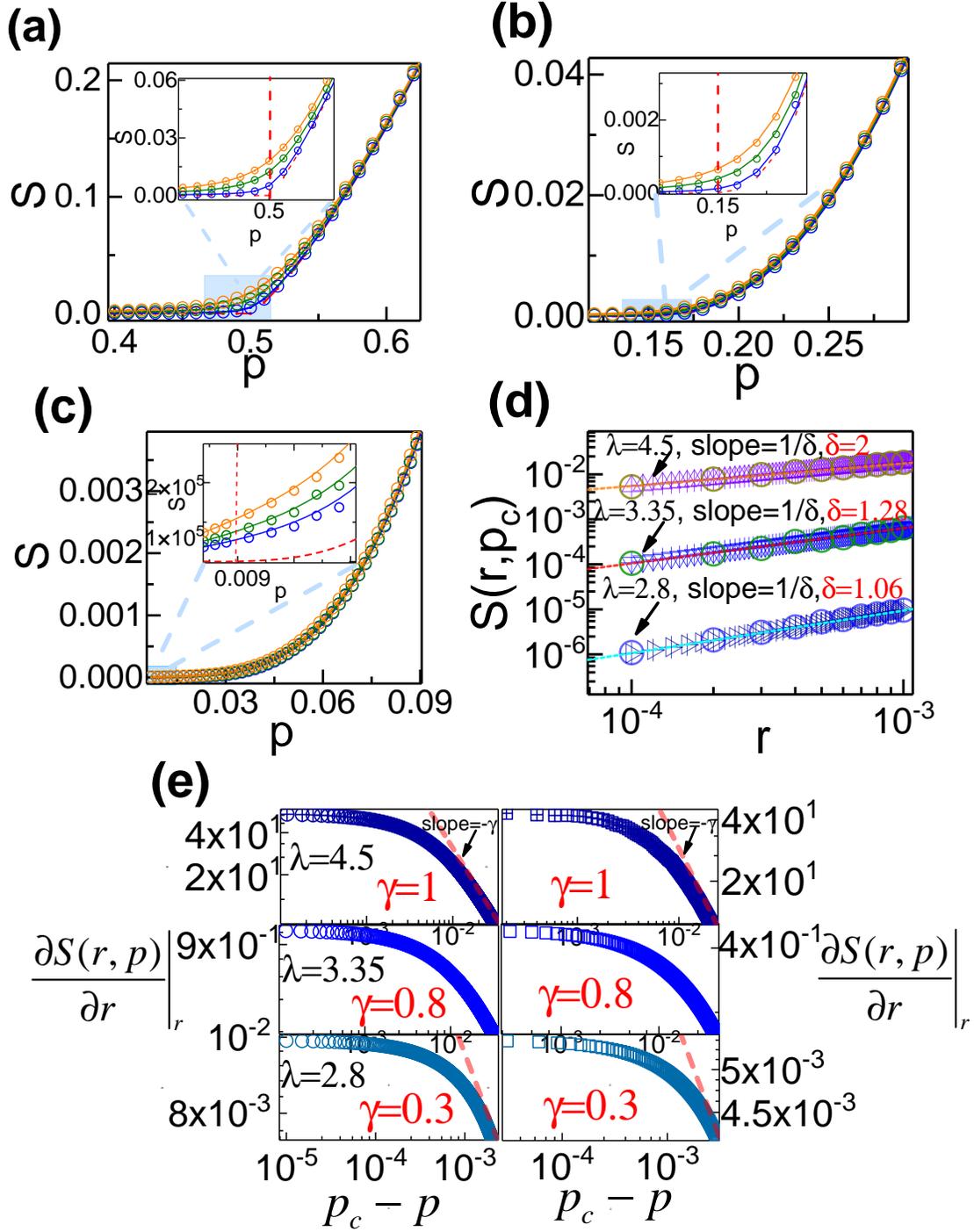}
\caption{\label{Fig:3} The size of the giant component $S(r,p)$ in SF networks as a function of $p$ for different $r$. (a) $\lambda=4.5$ with $r=0.0001$(blue), $r=0.0005$ (green) and $r=0.001$ (orange), (b) $\lambda=3.35$ for which $p_c\approx 0.149$ with $r=0.0001$(blue), $r=0.0005$ (green) and $r=0.001$ (orange), and (c) $\lambda=2.8$ with $r=0.0005$(blue), $r=0.0007$ (green) and $r=0.001$ (orange). Lines and symbols denote analytical and simulation results respectively, in which red line denote single SF network.(d) $S(r,p_{c})$ as a function of $r$ for different $\lambda$. Numerical and simulation results are denoted by circles and squares respectively. (e) $\frac{\partial S(r,p)}{\partial r}$ as a function of $p_{c} - p$ with $r=0.0001$ for $\lambda=4.5$, $r=0.0001$ for $\lambda=3.35$,
and  $r=0.0005$ for $\lambda=2.8$, left and right panels show numerical and simulation results. The simulation results were averaged over 1000 realizations with $k_{min}=2$, $k_{max}=10^{6}$, $N_{1}=N_{2}=10^{8}$ and $M_{inter}=N_{1}$.}
\end{figure}

By analyzing our analytical solution above,~\eqref{eq3}, we find that the $r$ interconnected nodes  have effects analogous to a magnetic field in a spin system. This is seen in that (i) for any non-zero fraction of interconnected nodes the system no longer undergoes a phase transition of the single module and (ii) field type critical exponents characterize the effect of $r$.
Fig.~\ref{Fig:2}(a) shows our analytic and simulation results for the giant component in two ER modules with average degree, $k=4$ and several $r$ values. It is seen that the percolation threshold is $p_c = 1/k = 1/4$ for a single ER module, however, the size of the giant component is above zero at $p_c$ when $r >0$. The theoretical and simulation results are in excellent agreement. 
Similar phenomenon is also observed for modules with a SF distribution with different values of $\lambda$, as seen in Fig.~\ref{Fig:3}(a).\\

We now investigate the scaling relations and critical exponents of our model, with $S(r,p)$, $p$ and $r$, serving as the analogues to magnetization, temperature, and external field respectively.
To quantify how the external field affects the percolation phase transition,
we define the critical exponents $\delta$, which relates the order parameter at the critical point to the magnitude of the field,
\begin{equation}\label{eq5}
S(r,p_c) \sim r^{1/\delta},
\end{equation}
and $\gamma$, which describes the susceptibility near criticality,
\begin{equation}\label{eq6}
\left (\frac {\partial S(r,p)} {\partial r}  \right)_{r\rightarrow 0} \sim  \left| p - p_c \right|^{-\gamma}.
\end{equation}

We begin by measuring $\delta$.
For ER modules, we obtain $\delta = 2$ from both theory and simulations [Fig.~\ref{Fig:2}(b)], which is the same as the known value for a mean-field random percolation exponent \cite{Stanley_1971}; For SF modules, we find that the value of $\delta$ varies with $\lambda$ as shown in Fig.~\ref{Fig:3}(d). When $\lambda >4$, the critical exponents are the expected mean-field values for regular percolation in infinite dimensions, and the universality class is the same as ER \cite{cohen_percolation_2002}. 
For $2<\lambda<3$, SF networks are known to undergo a transition only for $p\to 0$ and the critical exponents depend on $\lambda$. Similar results for different parameters are given in SI. For $3<\lambda<4$, it is known that $p_{c}>0$ and the critical exponents vary with $\lambda$ \cite{cohen_percolation_2002}. We find via simulations and analytically that $\delta$ changes with $\lambda$ as such: $\delta = 1.28$, for $\lambda = 3.35$ and  $\delta = 1.06$, for $\lambda = 2.8$.

We next consider the analogue of magnetic susceptibility, which has the scaling relation~\eqref{eq6}. Fig.~\ref{Fig:2}(c) presents the analytical (left) and simulations (right) results. For ER modules, we obtain $\gamma = 1$ for $p < p_c$, and also for $p > p_c$, see details in SI.
When considering SF modules, we find both in simulations and theory that $\gamma$ depends on $\lambda$, with
$\gamma = 1$ for $\lambda = 4.5$, $\gamma = 0.8$ for $\lambda = 3.35$, and
$\gamma = 0.3$ for $\lambda = 2.8$.

To test the scaling relations between the exponents, we note that for the single network ($m=1$), the order parameter follows
$S \sim (p - p_c)^{\beta}$ in the critical region with $\beta = 1$ for ER networks. For SF networks we know that $\beta = 1$ for  $\lambda >4$, $\beta = 1/(\lambda-3)$ for $3<\lambda <4$ and  $\beta = 1/(3 -\lambda)$ for $2<\lambda <3$ \cite{cohen_percolation_2002}. As shown below these values for $\beta$ fulfill together with the $\delta$ and $\gamma$ values found before the universal scaling relations that are well known in physical phase transitions \cite{Stanley_1971}.

The universality class of a system's phase transition is characterized by a set of critical exponents, since the various thermodynamic quantities are related, these critical exponents are not independent, but rather all exponents can be expressed in terms of only two exponents \emph{\textbf{\cite{Stanley_1971,domb2000phase}}}.
We find that this universal scaling hypothesis is also valid for our community model, both in ER and SF modules, based on the above values found for $\beta, \delta$ and $\gamma$.
Specifically, note that our values for these exponents are consistent with Widom's identity $\delta -1 = \gamma/\beta$ \cite{bunde2012fractals}.

%

%

\begin{figure}
\centering
\includegraphics[width=.9\linewidth]{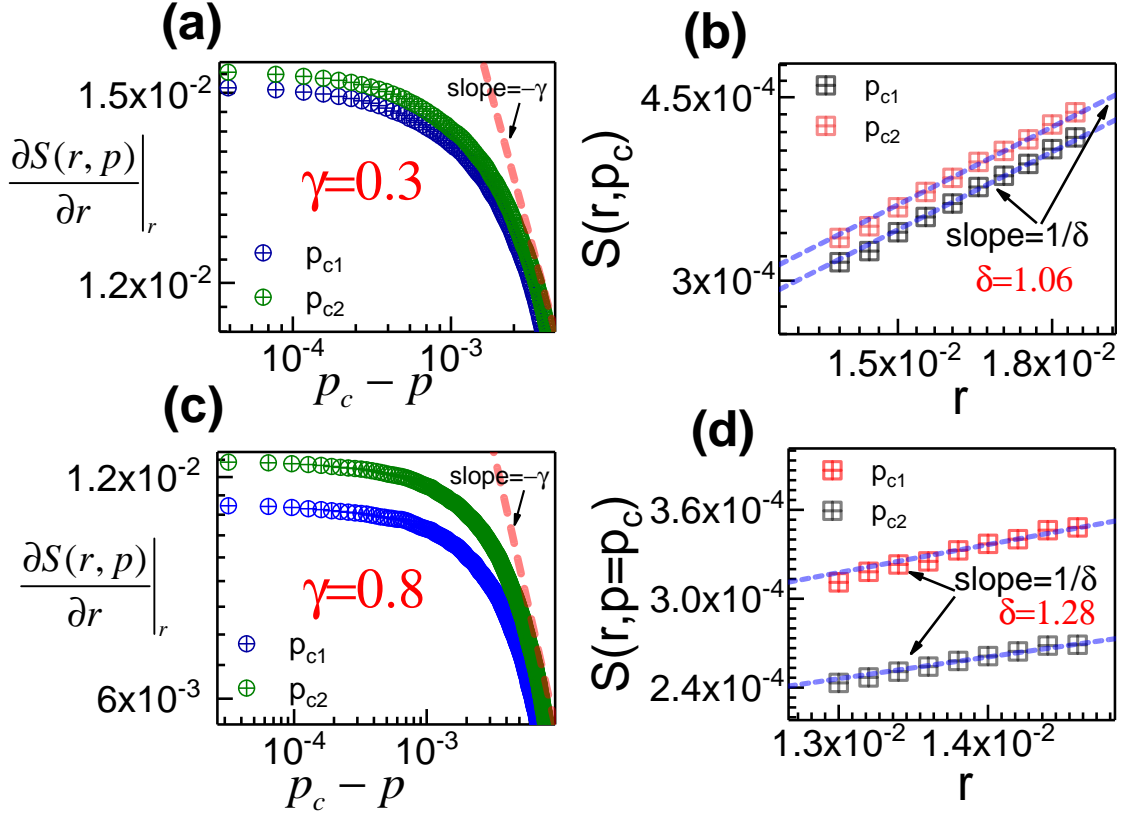}
\caption{\label{Fig:4} Critical scaling and exponents for two modules in each of two real world networks with $\lambda=2.8$ (a, b) and $\lambda=3.35$ (c, d). (a) $\frac{\partial S(r,p)}{\partial r}$ as a function of $p_{c} - p$ and (b) $S(r,p_{c})$ as a function of $r$ for the co-author dblp collaboration network.  (c) $\frac{\partial S(r,p)}{\partial r}$ as a function of $p_{c} - p$ and (d) $S(r,p_{c})$ as a function of $r$ for the co-author  MathSciNet collaboration network.  The parameters of each community and network
are summarized in SI. We average over 2000 realizations for each network.
%
}
\end{figure}

In the following, we test our framework on two real world examples: (i) the co-author collaboration network (DBLP) \cite{yang2015defining,networkrepository} and (ii) the co-authorship MathSciNet \cite{networkrepository,castellano2017relating,radicchi2015breaking} network built from the mathematical review collection of the American Mathematical Society [details in \textbf{Data and Methods}]. We use a greedy algorithm to
detect the community structure \cite{clauset2004finding},
and keep the largest two communities that have the same parameter $\lambda$ [the degree distribution for each community is given in SI].
 Fig.~\ref{Fig:4} shows the numerical results for modules of real networks with $\lambda = 2.8, 3.35$ respectively. We find that, the values of critical
exponents $\delta$ and $\gamma$ for the real networks are also consistent with theoretical results. One should note that
the percolation threshold in different communities is different in above real modules, as shown in SI.

\section{Discussion}
 In this work, we have introduced a network model of community structure and showed that the fraction of nodes with interconnections can be regarded as an external field such as in a physical phase transition. We solved the resilience of this system both numerically and analytically with excellent agreement. Our results show that a system becomes more stable and resilient as the fraction of nodes with interconnections increases. In particular, we find that the scaling relations governing the external field by defining critical exponents $\delta$ and $\gamma$ based on $S$, $p$ and $r$, are analogues to macroscopic magnetization, temperature and the external field respectively near criticality.  The values of the critical exponents are equivalent to the high dimensional values of magnetization transition in infinite dimensions for communities with a degree-distribution that is Poisson or SF with $\lambda>4$. For the case of SF degree distributions and $\lambda<4$ we find that $\delta$ and $\gamma$ depend on $\lambda$. Further, we find that these critical exponents obey the universal scaling relations known for physical systems near the phase transition. Similar results were found also for real social networks.

Our findings not only offer guidance on designing robust systems, but also make predictions about the nature of system failures (modeled through percolation). Our theory and model provide understanding of how to make the network more resilient by increasing the number of interconnected nodes as well as predicting its robustness.

In addition, we have extended percolation theory on networks by defining critical exponents for an external field. This work may also inspire further theoretical analysis and recognition of additional system properties that can be analogized as an external field. Although our theory is applied here to study the resilience of modules within a single network, it can be extended to study resilience of interdependent networks and multiplex networks.

\section{Data and Methods}
We applied the external field  model on two kinds of real collaboration networks: the co-author collaboration network (DBLP) \cite{yang2015defining,networkrepository} and the co-authorship MathSciNet network of mathematical review collections of the American Mathematical Society \cite{networkrepository,castellano2017relating,radicchi2015breaking}. In both networks, nodes are authors and an undirected edge between two authors exists if they have published at least one paper together. The community structure of the networks is detected by using a fast greedy algorithm \cite{clauset2004finding}. General information and statistical
features of these networks are summarized in SI.
To analyze the external field effects in real networks, we choose the largest two modules with the same scaling exponents. We then add non-duplicate interconnected links randomly among a fraction $r$ of nodes in both modules. The critical exponents of the external field for different $\lambda$ are analyzed in the critical region, as shown in Fig.~\ref{Fig:4}. And, the values of $p_c$ are determined by $S_\mathrm{cutoff}=0.0001$, where $S$ is smaller than or equal to $S_\mathrm{cutoff}$ for each individual module.

\section{acknowledgement}
We acknowledge the Israel-Italian collaborative
project NECST, the Israel Science Foundation, the Major Program of National Natural Science Foundation of China (Grant Nos. 71690242, 91546118), ONR, Japan
Science Foundation, BSF-NSF, and DTRA (Grant No.
HDTRA-1-10-1-0014) for financial support. This
work was partially supported by National Natural Science Foundation
of China (Grant Nos. 61403171, 71403105, 2015M581738 and 1501100B) and Key Research Program of Frontier Sciences, CAS, Grant No. QYZDJ-SSW-SYS019. J.F thanks the fellowship program funded by the Planning and Budgeting Committee of the Council for Higher Education of Israel. We thank Dr. Lucas Daniel Valdez for useful discussions.



\end{document}